\documentclass[runningheads]{llncs}

\usepackage[T1]{fontenc}
\usepackage{graphicx}
\usepackage{paralist}
\usepackage{amsmath,amssymb}
\usepackage{url}
\usepackage{hyperref}
\usepackage{microtype}
\usepackage{tikz}
\usetikzlibrary{positioning,shapes.misc}
\usetikzlibrary{decorations.pathreplacing}
\usetikzlibrary{decorations.pathmorphing}
\usetikzlibrary{arrows}
\usetikzlibrary{shapes}
\usepackage{subcaption}
\tikzset{
	mybrace/.style={decorate,decoration={brace,aspect=#1}}
}

\newcommand{\N}{\mathbb{N}}
\newcommand{\Z}{\mathbb{Z}}
\newcommand{\F}{\mathbb{F}}

\newtheorem{shortcoming}{Shortcoming}
\newtheorem{insight}{Insight}

\begin{document}
	
	\title{Insights Gained after a Decade of Cellular Automata-based Cryptography}
	
	\titlerunning{Insights after a Decade of CA-based Cryptography}
	
	\author{Luca Mariot\inst{1}\orcidID{0000-0003-3089-6517}}
	
	\authorrunning{L. Mariot}
	
	\institute{Semantics, Cybersecurity and Services Group, University of Twente \\
		Drienerlolaan 5, 7522NB Enschede, The Netherlands \\
		\email{l.mariot@utwente.nl}}
	
	\maketitle
	
	\begin{abstract}
		Cellular Automata (CA) have been extensively used to implement symmetric cryptographic primitives, such as pseudorandom number generators and S-boxes. However, most of the research in this field, except the very early works, seems to be published in non-cryptographic venues. This phenomenon poses a problem of relevance: are CA of any use to cryptographers nowadays? This paper provides insights into this question by briefly outlining the history of CA-based cryptography. In doing so, the paper identifies some shortcomings in the research addressing the design of symmetric primitives exclusively from a CA standpoint, alongside some recommendations for future research. Notably, the paper remarks that researchers working in CA and cryptography often tackle similar problems, albeit under different perspectives and terminologies. This observation indicates that there is still ample room for fruitful collaborations between the CA and cryptography communities in the future.
		
	\end{abstract}
	
	\section{Introduction}
	\label{sec:intro}
	Following the generic definition given by Goldreich~\cite{goldreich01}, the research field of \emph{cryptography} addresses the design of systems that should be resistant to any abuse. Historically, cryptography has been concerned mainly with secure communication: enabling two or more parties to communicate reliably in the presence of adversaries. In this sense, cryptography usually focuses on three aspects of secure communication: \emph{confidentiality}, meaning that only the intended recipient of a message can read it; \emph{integrity}, where the goal is to avoid any tampering of a message by an adversary; and \emph{authenticity}, ensuring that the legit sender has indeed transmitted the message.
	
	Nowadays, the confidentiality requirement of secure communication is still one of the primary use cases considered in cryptographic research, and it is mainly addressed through the design of appropriate \emph{encryption schemes}, usually classified as \emph{symmetric} or \emph{asymmetric} schemes. In what follows, we consider only the former breed of symmetric encryption scheme, where encryption and decryption use the same key.
	
	\emph{Cellular Automata} (CA) provide an interesting framework for symmetric encryption schemes. The motivation is twofold. First, by leveraging their massive parallelism, CA can realize cryptographic transformations very efficiently, especially when targeting hardware implementations. Second, the dynamic evolution of CA can become quite complex and unpredictable, depending on the underlying local rule used by the cells to update their states. In principle, cryptographic mappings can exploit this complexity so that they are not easily invertible without knowing the corresponding encryption key.
	
	CA started to raise some interest among cryptographers in the 1980s when Wolfram proposed a pseudorandom generator (PRG) based on the chaotic dynamics of a one-dimensional CA equipped with rule 30~\cite{wolfram85}. However, a few years later, Wolfram's generator was shown to be vulnerable against some attacks~\cite{meier-staff,koc-ca}, and this research trend apparently faded away in the cryptography community. Indeed, many subsequent works appeared in conferences and journals with a non-cryptographic focus~\cite{tomassini01,seredynski04,martin08,formenti14,leporati14,manzoni18}. This fact gives the impression that the CA and cryptography research communities diverged from each other, prompting a legitimate question: \emph{Are CA relevant for designing practical cryptographic primitives?}
	
	Actually, upon closer inspection of the literature, it turns out that CA \emph{are still} heavily used by cryptographers, with a large body of works published in cryptography-related venues~\cite{daemen94,panama,rijmen,liu,grassi22,giordani23}. The difference lies in the different terminology (e.g., shift-invariant mappings, rotation symmetric S-boxes, and liftings instead of CA) and the different approaches used to analyze the security and efficiency of the resulting primitives.
	
	This paper aims to investigate the above question through the following three-fold contribution:
	
	\begin{compactitem}
		\item Give a brief outline of the history of CA-based cryptography and how it evolved separately in the CA and cryptography research communities, highlighting commonalities and differences.
		\item List some shortcomings of the works on CA-based cryptography published in non-cryptographic venues, explaining why they are of limited utility.
		\item Provide some recommendations to mitigate the above shortcomings and align the research effort between the two communities of CA and cryptography.
	\end{compactitem}
	
	A critical remark is that the goal is not to give a comprehensive overview of CA-based cryptography. Instead, this paper revolves around two reference use cases: CA for the design of stream ciphers and block ciphers. These two research strands represent a reasonable subset of the literature, large enough to draw insights into CA's utility in cryptography. Consequently, the paper intentionally leaves out other cryptographic applications of CA such as hash functions~\cite{damgard89,daemen91,mihaljevic98}, public-key encryption schemes~\cite{kari92,clarridge09,applebaum10} and secret sharing schemes~\cite{rey05,ml-acri-2014,mgfl-desi-2020}. A second caveat is that, despite considering only CA for stream and block ciphers, the exposition of the paper is not entirely objective; rather, it builds upon the author's experience in this field accumulated in the past decade. Comments and suggestions on the perspectives proposed in this paper are encouraged and welcome.

	\section{Background}
	\label{sec:bg}
	This section quickly recalls the background concepts used throughout the paper, starting with the CA models most often used for cryptographic applications. Then, a recap of the basic notions and results related to symmetric cryptography is given, explicitly focusing on stream and block ciphers.
	
	\subsection{Cellular Automata}
	\label{subsec:ca}
	\emph{Cellular Automata} (CA) are a computational model defined over regular lattices of \emph{cells}. Each cell updates its state at each discrete time step by evaluating a \emph{local rule} on itself and its neighboring cells. The states of the cells range over a finite \emph{alphabet} $A$, and the lattice size can be either finite or infinite.
	
	Several variations of this model are possible depending on the size and dimension of the lattice, the topology of the neighborhood, the uniformity of the local rule, the alphabet, and so on. The simplest case corresponds to \emph{binary one-dimensional CA}: the cells are arranged over a one-dimensional array (finite or infinite), and each cell applies the same local rule to update its state in parallel. The alphabet is the binary set $A = \{0,1\}$, which, depending on the application, may be endowed with an appropriate algebraic structure such as a finite field (i.e., $A = \F_2$). Thus, the local rule is a \emph{Boolean function} $f: \{0,1\}^d \to \{0,1\}$, where $d$ is the \emph{diameter} of the CA and specifies the size of the neighborhood, that is, the number of cells that each cell needs to look at (including itself) to compute its next state.
	
	A further parameter is the \emph{offset} $\omega \in \{0, \cdots, d-1\}$, which defines the neighborhood's shift with respect to the cell under update. For example, a classic case is the \emph{symmetric neighborhood} where $d = 2r+1$ (with $r \in \N$ being called the \emph{radius} of the CA), and $\omega = r$. This means that the neighborhood of each cell is composed of itself, the $r$ cells to its left, and the $r$ cells to its right. Another typical setting is the \emph{one-sided neighborhood} with $\omega = 0$. In this case, each cell looks at itself and the $d-1$ cells to its right.
	
	When the cellular lattice is a bi-infinite string, the \emph{global rule} of a CA resulting from the parallel application of the local rule over all cells is a map $F$ from the \emph{full-shift space} $A^\Z$ to itself. The Curtis-Hedlund-Lyndon theorem~\cite{hedlund69} characterizes CA as those maps $F: A^\Z \to A^\Z$ that are both \emph{shift-invariant} and uniformly continuous under the Cantor distance. Research on infinite CA usually considers the long-term dynamic behavior of the dynamical system obtained by iterating the global map $F: A^\Z \to A^\Z$.
	
	Clearly, for practical applications (such as those concerning cryptography) CA are necessarily defined over finite lattices of $n \in N$ cells. This usually leads to the problem of updating the cells at the boundaries since they do not have a complete neighborhood. One common approach is to consider \emph{periodic boundary conditions}, where the one-dimensional lattice corresponds to a ring, with the first cell following the last one. This induces a global map $F:\{0,1\}^n \to \{0,1\}^n$ defined for all $x = (x_0,\cdots, x_{n-1}) \in \{0,1\}^n$ as:
	
	\begin{displaymath}
		F(x_0,\cdots,x_{n-1}) = (f(x_{0-\omega}, \cdots, x_{0-\omega+d-1}), \cdots f(x_{n-1-\omega}, \cdots, x_{n-\omega+d-2})) \enspace ,
	\end{displaymath}
	where all indices are taken modulo $n$. Therefore, a finite binary \emph{Periodic Boundary CA} (PBCA) corresponds to a vectorial Boolean function defined by \emph{shift-invariant coordinate functions}, with periodic boundary conditions. It is possible to iterate the global rule of a PBCA for an indefinite number of steps as in the infinite case. However, the dynamics are ultimately periodic, as after at most $2^n$ steps, the orbit of the CA will repeat itself.
	
	Periodic CA are probably the most used type of CA for cryptographic applications, ranging from the design of pseudorandom number generators\cite{wolfram85,formenti14,leporati14} to S-boxes~\cite{szaban08,keccak,picek17}. Another model considered to some extent in cryptography is the \emph{No-Boundary} CA (NBCA)~\cite{mariot19}. An NBCA only updates the cells with a complete neighborhood, meaning that the cells at the boundaries do not carry over to the next iteration, with the lattice shrinking after applying the global rule. Consequently, an NBCA can only be iterated for a finite number of time steps as long as there are enough cells to form at least one complete neighborhood. This is usually not a concern since certain applications (such as the design of S-boxes) only require a single evaluation of the global rule. Usually, the offset $\omega$ of NBCA is always $0$, leading to the following global rule $F: \{0,1\}^{n} \to \{0,1\}^{n-d+1}$:
	\begin{displaymath}
		F(x_0,\dots,x_{n-1}) = (f(x_0,\dots,x_{d-1}),f(x_1,\dots,x_{d}),\dots,f(x_{n-d},\dots,x_{n-1})) \enspace ,
	\end{displaymath}
	for all $x \in \{0,1\}^n$. As we mentioned earlier, in the binary case, the local rule of a CA is a Boolean function $f: \{0,1\}^d \to \{0,1\}$ of $d$ variables. The most natural way to represent $f$ is its truth table, which lists for each possible input vector $x \in \{0,1\}^d$ the corresponding output value $f(x)$. Assuming that the vectors of $\{0,1\}^d$ are totally ordered (e.g., through the lexicographic ordering), the $2^d$-bit vector $\Omega_f$ that represents the output column of the truth table uniquely identifies the rule $f$. The decimal encoding of this vector is also called the \emph{Wolfram code} of the rule, customary of the CA literature~\cite{wolfram83}. As an example, figure~\ref{fig:nbca-pbca} depicts respectively an NBCA and a PBCA with $n=6$ cells equipped with the elementary local rule 150 of diameter $d=3$, defined as $f(x_1, x_2, x_3) = x_1 \oplus x_2 \oplus x_3$.
	\begin{figure}[tb]
		\centering
		\begin{minipage}{0.5\textwidth}
			\centering
			\begin{tikzpicture}
				[->,auto,node distance=1.5cm, empt node/.style={font=\sffamily,inner
					sep=0pt}, rect
				node/.style={rectangle,draw,font=\bfseries,minimum size=0.5cm, inner
					sep=0pt, outer sep=0pt}]
				
				\node [empt node] (c)   {};
				\node [rect node] (c1) [right=0.1cm of c] {$1$};
				\node [rect node] (c2) [right=0cm of c1] {$0$};
				\node [rect node] (c3) [right=0cm of c2] {$0$};
				\node [rect node] (c4) [right=0cm of c3] {$1$};
				
				\node [empt node] (f1) [above=0.4cm of c2.east] {{\footnotesize
						$f(1,0,0) = 1$}};
				
				\node [rect node] (p2) [above=0.85cm of c1] {$0$};
				\node [rect node] (p1) [left=0cm of p2] {$1$};
				\node [rect node] (p3) [right=0cm of p2] {$0$};
				\node [rect node] (p4) [right=0cm of p3] {$0$};
				\node [rect node] (p5) [right=0cm of p4] {$0$};
				\node [rect node] (p6) [right=0cm of p5] {$1$};
				
				\node [empt node] (p7) [below=0.2cm of p1] {};
				\node [empt node] (p8) [right=0.07cm of p7] {};
				\node [empt node] (p12) [above=0.5cm of p1.east] {};
				\node [empt node] (p13) [above=0.5cm of p5.east] {};
				\node [empt node] (p14) [above=0.3cm of p13] {\phantom{M}};
				
				\draw [-, mybrace=0.25, decorate, decoration={brace,mirror,amplitude=5pt,raise=0.3cm}]
				(p1.west) -- (p3.east) node [midway,yshift=-0.3cm] {};
				\draw [-, draw=white, decorate, decoration={brace,amplitude=5pt,raise=0.3cm}]
				(p1.west) -- (p2.east) node [midway,yshift=0.3cm] {};
				\draw[->] (p8) -- (c1.north);
				\draw[->, draw=white] (p12) edge[bend left] (p13);
			\end{tikzpicture}
			\caption{NBCA}
		\end{minipage}%
		\begin{minipage}{0.5\textwidth}
			\centering
			\begin{tikzpicture}
				[->,auto,node distance=1.5cm, empt node/.style={font=\sffamily,inner
					sep=0pt}, rect
				node/.style={rectangle,draw,font=\sffamily\bfseries,minimum size=0.5cm, inner
					sep=0pt, outer sep=0pt}, grey node/.style={rectangle,draw,fill=gray!40,
					font=\sffamily\bfseries,minimum size=0.5cm, inner sep=0pt, outer sep=0pt}]
				
				\node [empt node] (c)   {};
				\node [rect node] (c1) [right=0.1cm of c] {$0$};
				\node [rect node] (c0) [left=0cm of c1] {$1$};
				\node [rect node] (c2) [right=0cm of c1] {$0$};
				\node [rect node] (c3) [right=0cm of c2] {$1$};
				\node [rect node] (c4) [right=0cm of c3] {$0$};
				\node [rect node] (c5) [right=0cm of c4] {$0$};
				
				\node [empt node] (f1) [above=0.2cm of c3] {{\footnotesize
						$f(1,1,0) = 0$}};
				
				\node [rect node] (p2) [above=0.85cm of c1] {$0$};
				\node [rect node] (p1) [left=0cm of p2] {$1$};
				\node [empt node] (p)  [left=0.1cm of p1] {};
				\node [rect node] (p3) [right=0cm of p2] {$0$};
				\node [rect node] (p4) [right=0cm of p3] {$0$};
				\node [rect node] (p5) [right=0cm of p4] {$0$};
				\node [rect node] (p6) [right=0cm of p5] {$1$};
				\node [grey node] (p7) [right=0cm of p6] {$1$};
				\node [grey node] (p8) [right=0cm of p7] {$0$};
				\node [empt node] (p9) [right=0.1cm of p8] {};
				
				\node [empt node] (p10) [below=0.2cm of p6] {};
				\node [empt node] (p11) [right=0.07cm of p10] {};
				\node [empt node] (p12) [above=0.5cm of p1.east] {};
				\node [empt node] (p13) [above=0.5cm of p7.east] {};
				
				\draw [-, mybrace=0.25, decorate, decoration={brace,mirror,amplitude=5pt,raise=0.3cm}]
				(p6.west) -- (p8.east) node [midway,yshift=-0.3cm] {};
				\draw [-, decorate, decoration={brace,amplitude=5pt,raise=0.3cm}]
				(p1.west) -- (p2.east) node [midway,yshift=0.3cm] {};
				\draw [-, decorate, decoration={brace,amplitude=5pt,raise=0.3cm}]
				(p7.west) -- (p8.east) node [midway,yshift=0.3cm] {};
				\draw[->] (p11) -- (c5.north);
				\draw[->] (p12) edge[bend left] (p13);
			\end{tikzpicture}
			\caption{PBCA}
		\end{minipage}
		\caption{Examples of NBCA and PBCA with local rule 150.}
		\label{fig:nbca-pbca}
	\end{figure}
	
	For a more comprehensive introduction to the basic notions and results related to CA, see the chapter by Kari~\cite{kari12}.
	
	\subsection{Cryptography}
	\label{subsec:crypto}
	This section focuses only on the \emph{confidentiality} of secure communication mentioned in the Introduction, which guarantees that only the intended recipient of a message can read it. Symmetric encryption schemes are one of the main approaches studied in cryptography to enforce this property. In an abstract setting, the sender, Alice, wants to transmit a plaintext message $P \in \{0,1\}^l$ to the receiver, Bob, over a channel wiretapped by an opponent, Oscar. The opponent can observe everything that passes over the channel. Therefore, Alice and Bob use an encryption scheme composed of the following steps. First, Alice gives the plaintext message $P$ in input to an encryption function $E$, parameterized over a secret key $K$. This key is known only to Alice and Bob, who agreed on it before the communication takes place\footnote{For instance, the key agreement can be achieved with public-key cryptography, which is, however, not the focus of this paper. Here, the assumption is that Alice and Bob already shared the encryption key securely.}. The output of the encryption function is a ciphertext $C \in \{0,\}^l$, which Alice transmits to Bob over the channel (and thus, eventually observed by Oscar). On the other end of the channel, Bob gives the ciphertext $C$ in input to a decryption function $D$,  parameterized on the same secret key $K$. The output of $D$ is the original plaintext message $P$ that Alice meant to send to Bob. The functions $E$ and $D$ must be the inverses of one another once they use the same secret key; otherwise, Bob cannot retrieve the correct message from the ciphertext.
	
	The confidentiality of this scheme relies on the assumption that Oscar cannot recover the plaintext message $P$ by observing the ciphertext $C$ when sent through the channel. Any sound encryption scheme should follow \emph{Kerchoff's principle}, according to which the security of the overall scheme should rely only on the secrecy of the key and not on the secrecy of the encryption and decryption functions, which are assumed to be public. Therefore, since $E$ and $D$ are known to Oscar, they must be designed not to leak any useful information on the plaintext if $K$ is not known.
	
	A standard classification of symmetric encryption schemes divides them into \emph{stream ciphers} and \emph{block ciphers}. The difference is that a stream cipher combines each plaintext symbol with a corresponding symbol of a \emph{keystream}, computed from the initial secret key through a \emph{keystream generator algorithm}. On the other hand, a block cipher encrypts the plaintext in \emph{blocks} of a fixed size, combining them iteratively with several \emph{round keys} generated from the secret key through a scheduling algorithm.
	
	One of the most well-known models of stream ciphers is the \emph{Vernam-like cipher}, in which the messages are binary strings of arbitrary length. The encryption amounts to the bitwise XOR between the plaintext and the keystream. Decryption is symmetric since by computing again the bitwise XOR between the ciphertext and the keystream, one obtains the original plaintext. The keystream generator algorithm is a \emph{Pseudorandom Generator} (PRG), which stretches the initial secret key into a pseudorandom sequence that matches the length of the plaintext.
	
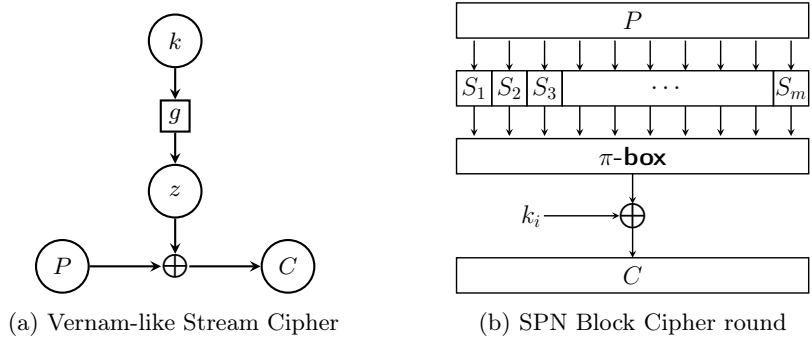
\begin{figure}[t]
		\centering
		\begin{subfigure}{.5\textwidth}
			\centering
			\begin{tikzpicture}
				[->,auto,node distance=1cm,
				circ node/.style={circle,thick,draw,font=\sffamily\bfseries, minimum size = 0.7cm},
				empt node/.style={font=\sffamily,inner sep=0pt,minimum size=0pt},
				rect node/.style={rectangle,thick,draw,font=\sffamily\bfseries}]
				
				\node[circ node] (1) {$k$};
				\node[rect node] (2) [below of=1] {$g$};
				\node[circ node] (3) [below of=2] {$z$};
				\node[empt node] (4) [below of=3] {$\bigoplus$};
				\node[circ node] (5) [left of=4, node distance = 1.5cm] {$P$};
				\node[circ node] (6) [right of=4, node distance = 1.5cm] {$C$};
				
				\draw [->, thick, shorten >=0pt,shorten <=0pt,>=stealth] (1) -- (2);
				\draw [->, thick, shorten >=0pt,shorten <=0pt,>=stealth] (2) -- (3);
				\draw [->, thick, shorten >=0pt,shorten <=0pt,>=stealth] (3) -- (4);
				\draw [->, thick, shorten >=0pt,shorten <=0pt,>=stealth] (5) -- (4);
				\draw [->, thick, shorten >=0pt,shorten <=0pt,>=stealth] (4) -- (6);
			\end{tikzpicture}
			\caption{\small{Vernam-like Stream Cipher}}
			\label{fig:stream}
		\end{subfigure}%
		\begin{subfigure}{.5\textwidth}
			\centering
			\resizebox{.8\textwidth}{!}{
				\Large
				\begin{tikzpicture}
					[->,auto,node distance=1.5cm, empt node/.style={font=\sffamily,inner
						sep=0pt, outer sep=-0.5pt}, rect
					node/.style={rectangle,thick,draw,font=\sffamily\bfseries,minimum size=0.7cm, inner
						sep=0pt, outer sep=0pt}, rect1
					node/.style={rectangle,thick,draw=white,font=\sffamily\bfseries,minimum size=0.7cm, inner sep=0pt, outer sep=0pt}]
					
					\node [rect node] (pt) [minimum width=7cm] {$P$};
					\node [empt node] (e)  [below=1cm of pt] {};
					\node [rect1 node] (s5) [left=0cm of e] {};
					\node [rect1 node] (s4) [left=0cm of s5] {};
					\node [rect node] (s3) [left=0cm of s4] {$S_3$};
					\node [rect node] (s2) [left=0cm of s3] {$S_2$};
					\node [rect node] (s1) [left=0cm of s2] {$S_1$};
					\node [rect1 node] (s6) [right=0cm of e] {};
					\node [rect1 node] (s7) [right=0cm of s6] {};
					\node [rect1 node] (s8) [right=0cm of s7] {};
					\node [rect1 node] (s9) [right=0cm of s8] {};
					\node [rect node] (s10) [right=0cm of s9] {$S_{m}$};
					\node [rect node] (r) [right=0cm of s3, minimum width=4.2cm] {$\cdots$};
					
					\node [empt node] (e1) [above=0.6cm of s1] {};
					\node [empt node] (e2) [above=0.6cm of s2] {};
					\node [empt node] (e3) [above=0.6cm of s3] {};
					\node [empt node] (e4) [above=0.6cm of s4] {};
					\node [empt node] (e5) [above=0.6cm of s5] {};
					\node [empt node] (e6) [above=0.6cm of s6] {};
					\node [empt node] (e7) [above=0.6cm of s7] {};
					\node [empt node] (e8) [above=0.6cm of s8] {};
					\node [empt node] (e9) [above=0.6cm of s9] {};
					\node [empt node] (e10) [above=0.6cm of s10] {};
					
					\node [empt node] (b1) [below=0.6cm of s1] {};
					\node [empt node] (b2) [below=0.6cm of s2] {};
					\node [empt node] (b3) [below=0.6cm of s3] {};
					\node [empt node] (b4) [below=0.6cm of s4] {};
					\node [empt node] (b5) [below=0.6cm of s5] {};
					\node [empt node] (b6) [below=0.6cm of s6] {};
					\node [empt node] (b7) [below=0.6cm of s7] {};
					\node [empt node] (b8) [below=0.6cm of s8] {};
					\node [empt node] (b9) [below=0.6cm of s9] {};
					\node [empt node] (b10) [below=0.6cm of s10] {};
					
					\node [rect node] (pbox) [below=2cm of pt, minimum width=7cm] {$\pi$-box};
					\node [empt node] (xor) [below=0.6cm of pbox] {$\bigoplus$};
					\node [empt node] (k) [left=1.5cm of xor, inner sep=2pt] {$k_i$};
					\node [rect node] (ct) [below=0.6cm of xor, minimum width=7cm] {$C$};
					
					\draw[->,thick, shorten >=0pt,shorten <=0pt,>=stealth] (pbox) -- (xor);
					\draw[->,thick, shorten >=0pt,shorten <=0pt,>=stealth] (xor) -- (ct);
					\draw[->,thick, shorten >=0pt,shorten <=0pt,>=stealth] (k) -- (xor);
					\draw[->,thick, shorten >=0pt,shorten <=0pt,>=stealth] (e1) -- (s1);
					\draw[->,thick, shorten >=0pt,shorten <=0pt,>=stealth] (e2) -- (s2);
					\draw[->,thick, shorten >=0pt,shorten <=0pt,>=stealth] (e3) -- (s3);
					\draw[->,thick, shorten >=0pt,shorten <=0pt,>=stealth] (e4) -- (s4);
					\draw[->,thick, shorten >=0pt,shorten <=0pt,>=stealth] (e5) -- (s5);
					\draw[->,thick, shorten >=0pt,shorten <=0pt,>=stealth] (e6) -- (s6);
					\draw[->,thick, shorten >=0pt,shorten <=0pt,>=stealth] (e7) -- (s7);
					\draw[->,thick, shorten >=0pt,shorten <=0pt,>=stealth] (e8) -- (s8);
					\draw[->,thick, shorten >=0pt,shorten <=0pt,>=stealth] (e9) -- (s9);
					\draw[->,thick, shorten >=0pt,shorten <=0pt,>=stealth] (e10) -- (s10);
					
					\draw[->,thick, shorten >=0pt,shorten <=0pt,>=stealth] (s1) -- (b1);
					\draw[->,thick, shorten >=0pt,shorten <=0pt,>=stealth] (s2) -- (b2);
					\draw[->,thick, shorten >=0pt,shorten <=0pt,>=stealth] (s3) -- (b3);
					\draw[->,thick, shorten >=0pt,shorten <=0pt,>=stealth] (s4) -- (b4);
					\draw[->,thick, shorten >=0pt,shorten <=0pt,>=stealth] (s5) -- (b5);
					\draw[->,thick, shorten >=0pt,shorten <=0pt,>=stealth] (s6) -- (b6);
					\draw[->,thick, shorten >=0pt,shorten <=0pt,>=stealth] (s7) -- (b7);
					\draw[->,thick, shorten >=0pt,shorten <=0pt,>=stealth] (s8) -- (b8);
					\draw[->,thick, shorten >=0pt,shorten <=0pt,>=stealth] (s9) -- (b9);
					\draw[->,thick, shorten >=0pt,shorten <=0pt,>=stealth] (s10) -- (b10);
				\end{tikzpicture}
			}
			\caption{\small{SPN Block Cipher round}}
			\label{fig:block}
		\end{subfigure}%
		\caption{Diagrams for Vernam-like stream ciphers and SPN block ciphers.}
		\label{fig:scdiag}
	\end{figure}

	The \emph{Substitution-Permutation Network} (SPN) is a typical design paradigm for block ciphers. Assuming that the plaintext is divided into equal-sized blocks, each block is encrypted by first applying a \emph{confusion phase}, followed by a \emph{diffusion phase}, and finally by the \emph{key combination phase} with the current round key. This procedure is then iterated for a certain number of \emph{rounds}. Decryption performs the same operations but in reverse order. Figure~\ref{fig:scdiag} displays the block diagrams for the Vernam-like stream cipher and the SPN block cipher.
	
	Shannon~\cite{shannon49} set forth the properties of \emph{confusion} and \emph{diffusion} as guiding principles for the design of secure symmetric encryption schemes to withstand statistical attacks. Confusion aims to make the relationship between the plaintext message and the secret key as complex as possible. On the other hand, diffusion aims to spread the statistical structure of the plaintext over the ciphertext, the idea being that the ciphertext should look indistinguishable from random noise.
	
	Per se, the notions of confusion and diffusion are abstract and need to be specified concretely with respect to the underlying cipher model. For example, the security of a Vernam-like stream cipher can be reduced to the properties of the underlying PRG, which are, in turn, designed by following different approaches. In the \emph{combiner model}, the PRG comprises $n$ Linear Feedback Shift Registers (LFSRs) initialized with the secret key. At each clock cycle, a Boolean function $f: \F_2^n \to \F_2$ of $n$ variables combines the outputs of the leftmost LFSRs bits. The output of $f$ constitutes the next bit in the pseudorandom keystream. The security of this PRG is then analyzed in terms of the properties of the combiner function $f$. The rationale is that some cryptanalytic attacks can be carried out more efficiently by an adversary if $f$ does not meet specific criteria. The goal of these attacks could include, for instance, the recovery of the LFSR initial states (and thus, the secret key).
	
	The study of the cryptographic criteria of Boolean functions is a whole research field in itself, and a proper overview would be out of scope for the present paper. Carlet's book~\cite{carlet21} is the standard reference for this topic. Essential properties recalled in the remainder of this paper include \emph{balancedness} (the truth table vector of $f$ is composed of an equal number of zeros and ones), \emph{nonlinearity} (the Hamming distance of $f$ from the set of linear functions, which are defined using only XOR operations), \emph{correlation immunity} (the statistical correlation between subsets of input variables and the output of $f$) and \emph{algebraic degree} (the degree of the polynomial that defines the \emph{Algebraic Normal Form} of $f$).
	
	A similar discussion applies to block ciphers designed around the SPN paradigm, whose security primarily resides in the properties of the confusion and diffusion layers. The diffusion layer is usually implemented as a transformation (a P-box) that acts globally on the whole cipher state to shuffle the message bits. The confusion layer, on the other hand, is usually realized through a set of smaller \emph{Substitution boxes} (S-Boxes), which act on smaller chunks of the block. From a mathematical standpoint, S-boxes are basically \emph{vectorial} Boolean functions $S: \F_2^n \to \F_2^n$ with $n$ inputs and $n$ outputs. Analogously to the combiner model for stream ciphers, the security of the confusion layer can be reduced to the study of the security properties of the S-boxes. Similarly to the one-bit output case, criteria of interest include \emph{balancedness} (which here corresponds to bijectivity, necessary to allow for decryption), nonlinearity, and \emph{differential uniformity} (which is related to differential cryptanalysis attacks). Further information on the properties of S-boxes can be found in~\cite{carlet21}. An excellent reference for a general introduction to cryptographic schemes and models is~\cite{lindell21}.
	
	\section{Stream Ciphers Based on CA}
	\label{sec:stream}
	This section reviews the history of CA-based cryptography regarding the design of pseudorandom number generators and stream ciphers. Next, it gives an overview of the primary attacks on this approach and outlines some improvements in subsequent research. Finally, the section concludes with a discussion of the pitfalls in the CA literature on the design of stream ciphers and gives some insights on how to overcome them.
	
	\subsection{Early Works}
	\label{subsec:early}
	
	The first attempt to define a stream cipher based on CA can be traced back to Wolfram~\cite{wolfram85}. The author proposed using an \emph{elementary} CA (i.e., a one-dimensional CA with diameter $d=3$ and symmetric neighborhood) equipped with the local rule 30. The underlying idea is to represent the stream cipher's secret key as the CA lattice's initial configuration, with periodic boundary conditions. Then, the CA evolves for multiple time steps, and the sequence of states taken by the central cell is the pseudorandom sequence of a Vernam-like stream cipher. 
	
	The rationale for choosing rule 30 was that Wolfram classified it among the so-called \emph{class 3} rules~\cite{wolfram84}, which induce CA with chaotic dynamics. Hence, it is reasonable to expect that CA based on such rules exhibit good confusion and diffusion properties. Wolfram thoroughly investigated the dynamics of his system based on rule 30 using statistical and empirical tests, such as the computation of Lyapunov's exponent. The results of these tests indicated that a CA equipped with this local rule could produce pseudorandom sequences of good quality that are difficult to predict for an adversary when used in a Vernam-like stream cipher. Wolfram proposed using a CA with a lattice size of at least $n=127$ cells, adopting different sampling strategies to destroy correlations among the sequence of bits. One of these strategies was to sample the central cell in alternating time steps. This strategy requires iterating the CA for twice the number of steps to construct a sufficiently long pseudorandom sequence.
	
	\subsection{Attacks on Wolfram's PRNG and Improvements}
	\label{subsec:attacks}
	It did not take long until a few attacks were demonstrated against Wolfram's PRG, proving that it was very weak from a cryptographic
	point of view. Meier and Staffelbach proposed the first attack in~\cite{meier-staff}, where the authors showed that a CA equipped with rule 30 is vulnerable to a known-plaintext attack based on the algebraic structure of rule 30. More precisely, the algebraic normal form of rule 30 is $f(x_1, x_2, x_3) = x_1 \oplus x_2 \oplus x_1x_2 \oplus x_3$; thus, it depends linearly on the rightmost variable $x_3$. This property is also called \emph{permutivity} or \emph{quasi-linearity}~\cite{m-physicad-1997} and allows one to rewrite the rule in an equivalent expression where the initial seeds are not equiprobable. Later, Koc and Apohan~\cite{koc-ca} proposed a second attack on Wolfram's PRG, exploiting the best affine approximation of rule $30$.
	
	The two attacks described above are a direct consequence of the poor cryptographic properties of rule 30 when interpreted as a Boolean function. More specifically, Meier and Staffelbach's attack can be carried out efficiently because rule 30 does not satisfy first-order correlation immunity. In contrast, Koc and Apohan's attack depends on the fact that the nonlinearity of rule 30 is very low. Therefore, a possible solution would be to select a different local rule to mitigate these issues. Martin~\cite{martin08} showed by an exhaustive search that no elementary local rules are simulatenously nonlinear and first-order correlation immune.
	
	Thus, the only way to salvage Wolfram's PRG is to look for local rules of a larger diameter with a better trade-off of cryptographic properties, which posits a combinatorial optimization problem. The space of $d$-variable Boolean functions $f: \F_2^d \to \F_2$ contains $2^{2^d}$ elements; hence it grows super-exponentially in $d$. Exhaustive enumeration is possible only up to $d=5$ variables. For larger diameters, one must resort to algebraic constructions~\cite{carlet21}, metaheuristic optimization algorithms~\cite{djurasevic23}, or combinatorial techniques to reduce the search space. Formenti et al. undertook the latter path in~\cite{formenti14}. There, the authors used the classification of Boolean functions of $d=5$ variables conducted by Braeken et al.~\cite{braeken05}, which partitions the space into 48 equivalence classes. Formenti et al. then performed an exhaustive search among representatives of these classes, finding those susceptible to preserve first-order correlation immunity after two iterations of a CA. Finally, the authors performed statistical tests from the DIEHARD suite~\cite{marsaglia08} on the pseudorandom sequences produced by Wolfram's PRG with these new rules plugged in, sifting those that passed all the tests.
	
	Leporati and Mariot~\cite{leporati13} also focused on the space of rules of diameter $d=5$. In this case, the authors considered \emph{bipermutive local rules} (i.e., rules that depend linearly on both the left and rightmost variables). The rationale was that such rules induce CA that satisfy strong definitions of topological chaos~\cite{cattaneo00}. Therefore, they seem good candidates for generating pseudorandom sequences with good statistical properties. Moreover, Leporati and Mariot proved that bipermutive rules are $1$-resilient, i.e., balanced and first-order correlation immune\footnote{This is actually a particular case of a result proved much earlier by Siegenthaler~\cite{sieg-ci}}. Since the space of bipermutive local rules of diameter $5$ is composed of 256 rules, the authors performed an exhaustive search to find those rules that are both 2-resilient and have maximum nonlinearity. Subsequently, the authors used statistical tests from the ENT~\cite{ent-test} and NIST~\cite{nist-test} test suites to filter only those rules that produced good pseudorandom sequences when plugged into Wolfram's PRG. The same authors later expanded this research in~\cite{leporati14} by considering further properties besides nonlinearity and resiliency and extended the exhaustive search experiments for bipermutive local rules of diameter $d=7$.
	
	More recently, Manzoni et al.~\cite{manzoni18} considered Wolfram's PRG from a different angle, namely by extending the model to \emph{asynchronous CA}, where the cells do not update necessarily in parallel at each time step. An interesting finding from the statistical tests of the NIST suite reported in~\cite{manzoni18} is that certain local rules produce better pseudorandom sequences under a small amount of asynchrony.
	
	\subsection{Shortcomings and Insights}
	\label{subsec:ins-stream}
	The story of Wolfram's PRG shows quite clearly the first shortcoming shared by many works in the literature of CA-based cryptography:
	\begin{shortcoming}
		\label{shc:first}
		Grounding security claims of CA-based cryptographic primitives on statistical or empirical tests or criteria unrelated to cryptography (e.g., chaos-based properties) can be misleading.
	\end{shortcoming}
	Rule 30 is a paradigmatic example of this pitfall: despite faring well under statistical tests, and even if it induces CA that are chaotic in Devaney's sense~\cite{cattaneo00}, it is entirely useless to generate pseudorandom sequences for cryptographic purposes, as demonstrated by the Meier-Staffelbach and Koc-Apohan attacks. The underlying problem here is that passing a battery of statistical tests is only a \emph{necessary condition} for a good pseudorandom generator. As such, these tests help discard bad generators, but it is not possible to claim security just by observing that a rule is passing all tests in a suite. The reason is that there is an infinity of statistical tests, and passing a finite subset of them does not guarantee that the PRG will not fail on others. Similarly, system-theoretic and chaos-related properties are not necessarily related to actual cryptanalytic attacks. Although they might tell us that a CA is difficult to predict or invert by just observing its current state, cryptanalysis usually considers more powerful attack models, where the adversary can also have access to some pairs of plaintexts and ciphertexts (as in a known-plaintext attack). For this reason, the cryptographic properties studied for Boolean functions target specific attacks; this leads to the following first insight to deal with this problem of CA-based cryptography:
	\begin{insight}
		\label{ins:first}
		Statistical tests are fine only to filter out bad CA-based cryptographic primitives. For a proper security analysis, at a minimum, the cryptographic properties of the underlying local rules should be carefully investigated.
	\end{insight}
	Under this light, even recent works such as~\cite{manzoni18} should be revisited to check whether the selected rules have good cryptographic properties. 
	
	Is Insight~\ref{ins:first} enough? If so, the design of good CA-based pseudorandom generators for stream ciphers could be reduced just to the search of local rules with a good trade-off of cryptographic properties along the lines of works such as~\cite{formenti14,leporati14}. However, as we mentioned in Section~\ref{subsec:crypto}, the criteria for Boolean functions refer to specific models of keystream generators, e.g., the combiner model. This is quite different from Wolfram's PRG model, where we are iterating the CA for multiple time steps and sampling the trace of the central cell. The \emph{filter model}~\cite{carlet21} looks more similar to Wolfram's PRG but still has notable differences. Consequently, this leads to the following shortcoming:
	\begin{shortcoming}
		\label{shc:second}
		Claiming that a Wolfram-like PRG is secure due to the cryptographic properties of the underlying local rule is not enough, because some attacks on the combiner or filter model might not be relevant in the CA setting. On the other hand, it could be that the current known cryptographic criteria of Boolean functions do not capture other types of attacks unique to the CA model.
	\end{shortcoming}
	The remark above is also probably connected to why research on CA-based stream ciphers is seldom published in cryptographic venues. After the initial attempt by Wolfram (whose rule 30-based PRG was published in CRYPTO), cryptographers preferred to shift to other designs where attacks can be modeled and investigated more easily. Wolfram's PRG model still needs to be better understood regarding cryptanalytic attacks. The Meier-Staffelbach and Koc-Apohan attacks concern the correlation immunity and the nonlinearity of the CA local rule, respectively. However, there are many other criteria of Boolean functions tailored for other attacks on the combiner and the filter model. This remark brings up the following:
	\begin{insight}
		\label{ins:second}
		Consistently link the proposed CA model with the security properties and the related attacks. Each cryptographic property considered in the security analysis should be thoroughly motivated in terms of attacks specifically tailored for the CA model, not the combiner or the filter model.
	\end{insight}
	To provide a concrete research direction for the future, the \emph{algebraic degree} is among those properties of Boolean functions that should be considered in the works on CA-based PRG. This property is, in fact, quite crucial since it is related to the Berlekamp-Massey attack in the combiner model~\cite{massey69} and the R{\o}njom-Helleseth algebraic attack in the filter model~\cite{ronjom07}.
	
	The upshot of these attacks is that the algebraic degree of the underlying Boolean function should be as high as possible (ideally $n-1$, which is the maximum possible for balanced functions). A related cryptographic criterion, the \emph{algebraic immunity} of $f$, should also be the highest possible to thwart the R{\o}njom-Helleseth algebraic attack. Moreover, especially in the combiner model, one should use combiner functions of at least $n=13$ variables.
	
	It would be interesting to investigate the algebraic degree and the algebraic immunity of the local rules used in CA-based PRG. This direction would entail, in particular, adapting the Berlekamp-Massey and the R{\o}njom-Helleseth attack to the CA model, provided they are relevant. If that is the case, it would also be interesting to consider the minimum diameter a local rule should possess to make these attacks inefficient. CA are usually attractive because even small local rules can generate CA with complex behaviors. Efficiency issues would likely arise in a CA-based PRG model when considering large diameters such as $d=13$.
	
	As remarked by Carlet~\cite{carlet21}, the combiner and the filter models are only models and are mainly helpful in studying attacks. However, the designs of stream ciphers are more complicated in practice. Hence, a practical stream cipher based on CA would likely require a design that is more sophisticated than Wolfram's PRG. As a concrete research direction, Mariot et al.~\cite{mariot17} observed that the {\sc Grain} stream cipher~\cite{hell08}, selected as a finalist for the eSTREAM competition, has a structure that closely resembles the computation of a CA preimage. In particular, the state of {\sc Grain} is determined by concatenating a Nonlinear Feedback Shift Register (NFSR) with a LFSR. An interesting open problem is determining whether the tools developed in~\cite{mariot17} could be exploited to cryptanalyze the {\sc Grain} stream cipher or develop analogous models for concrete stream ciphers.
	
	\section{Block Ciphers Based on CA}
	\label{sec:block}
	This section surveys the main results related to the design of CA-based symmetric primitives for block ciphers. Researchers in the CA and cryptography communities have extensively considered this use case, although it does not seem to be the case from a superficial view. The reason is that researchers in symmetric cryptography adopt a different terminology to refer to cellular automata, calling them instead shift-invariant mappings, rotation symmetric S-boxes, or liftings. The following discussion attempts to identify commonalities between the two research tracks, drawing related shortcomings and insights at the end.     
	
	\subsection{Encryption by Iterating CA}
	\label{subsec:iter}
	Historically, the research concerning block ciphers based on CA followed an approach similar to the stream cipher/PRG strand initiated by Wolfram in~\cite{wolfram85}: the basic idea is to exploit the complexity of the patterns emerging from the iterated evolution of the CA for multiple time steps, considered as a dynamical system. With block ciphers, new issues arise that are not relevant for CA-based PRGs: the most evident one is that \emph{reversibility} of the underlying CA is often sought since it is necessary for decryption. On the other hand, reversibility is already guaranteed by the symmetry of the XOR operation in Vernam-like stream cipher, so it is not a primary concern in CA-based PRGs\footnote{Actually, (partial) reversibility could even represent a security problem, as demonstrated by the Meier-Staffelbach attack on Wolfram's PRG.}. 
	
	Gutowitz~\cite{gutowitz93} was the first to propose a block cipher wholly based on the dynamics of CA. The cipher design employed both irreversible and reversible CA, respectively, for diffusion and confusion. The irreversible part consisted of permutive CA, for which a simple algorithm exists to compute a random preimage of a configuration. On the other hand, the author proposed to employ \emph{block CA} for the reversible part, where a fixed mapping is applied on sub-blocks of the configuration, shifted one cell to the right with periodic boundary conditions to emulate the shift-invariance property of CA. The reversibility of this system is trivially guaranteed by the fact that the mapping is a permutation.
	
	Seredynsky et al.~\cite{seredynski04} investigated \emph{second-order} CA for the design of S-boxes, where reversibility is enforced by computing the XOR between the output of the local rule with the state of the cell at the \emph{previous} time step. The authors investigated the S-boxes defined by iterating second-order CA for multiple time steps with respect to their \emph{avalanche effect}, a property related to the resistance of block ciphers against differential cryptanalysis~\cite{webster85}.
	
	Marconi et al. investigated a third approach by proposing \texttt{Crystal}~\cite{marconi06}, a block cipher based on the dynamics of \emph{Lattice Gases Automata} (LGA). LGA are a CA variant usually employed as a discrete fluid model, based on the \emph{collision-propagation} paradigm. In particular, if the local rule implements a reversible physical process, the overall system is reversible and equal to its inverse. This observation is especially interesting for constrained hardware implementations since the same circuitry can be used both for encryption and decryption. The authors claimed that \texttt{Crystal} can provide scalable and efficient encryption (up to 10Gbps on dedicated hardware). However, the security of the block cipher was analyzed only through a few empirical tests.   
	
	Yet another direction was explored by Szaban et al., who designed S-boxes based on the iteration of CA~\cite{szaban08}. Instead of starting from known reversible CA, the authors considered all elementary local rules, retaining only those that resulted in invertible S-boxes of size $8\times 8$ with the best nonlinearity and autocorrelation properties after evolving the CA for a certain number of steps. More recently, Ghoshal et al.~\cite{ghoshal18} investigated S-boxes of size $4\times 4$ with optimal nonlinearity and differential uniformity defined by multiple iterations of CA rules and showed efficient \emph{threshold implementations} for them.
	
	\subsection{The Single-Step CA Approach}
	\label{subsec:single}
	All works surveyed in the previous section exploit the iterated behavior of CA to implement a block cipher or a lower-level primitive thereof. A different perspective is considering a CA evolved only for a single time step instead. This approach was pioneered by Daemen et al. in~\cite{daemen94}, where the authors investigated CA with local rules defined by \emph{complementing landscapes}. A complementing landscape is a simple regular expression representing a set of patterns occurring in the neighborhood of a cell. When any of these patterns appear, the cell flips its state; otherwise, it remains in the same state. Daemen et al. considered a simple local rule, which they named $\chi$, that flips the bit of the cell if the pattern $10$ occurs in the two neighboring cells to its right\footnote{$\chi$ is actually the elementary rule $210$ under Wolfram's code convention, taken however with offset $\omega = 0$.}. Interestingly, the authors showed that the resulting PBCA is invertible only if the lattice size is odd due to an inversion algorithm based on the idea of \emph{seeds} and \emph{leaps}. However, this inversion algorithm is inherently sequential, which means that the inverse is not described by a cellular automaton.
	
	Additionally, the authors of~\cite{daemen94} remarked that $\chi$ has good correlation and propagation characteristics, making it a good candidate to construct S-boxes and confusion layers resistant to linear and differential cryptanalysis. This rule (and its one's complement, called $\gamma$) appeared in the design of several primitives, such as the hash functions {\sc Panama}~\cite{panama} and {\sc RadioGat\'{u}n}~\cite{radiogatun}. Interestingly, the permutation in the {\sc Keccak} sponge construction~\cite{radiogatun}, an instantiation of which has been adopted by the NIST as the SHA-3 standard for cryptographic hash functions, uses a PBCA of size $n=5$ equipped with rule $\chi$ as its sole nonlinear component. This is perhaps the best-known example of a CA-based cryptographic primitive that partakes in the design of a major cryptographic standard, although it is pretty well hidden. Indeed, the authors of {\sc Keccak} do not call this mapping a CA but rather a shift-invariant function.
	
	Daemen further investigated the idea of applying complementing landscapes CA to design symmetric ciphers in his PhD thesis~\cite{daemen95}. There, the author also considered the so-called \emph{locally invertible} transformations: CA whose local rules are still described by a set of complementing landscapes, where, however, the inverse is itself a CA. In this case, the CA is actually an involution, i.e., the local rule equals its inverse. As noted in the previous section with the LGA paradigm, having an involution is interesting for implementation reasons since the same operation can be used both for encryption and decryption. However, very recently, Mariot et al.~\cite{mariot21} showed through evolutionary algorithms that locally invertible CA usually have bad cryptographic properties. The reason is that the complementing landscapes must not overlap in this case; otherwise, reversibility is destroyed. This constraint forces the cell to flip its state only rarely. Consequently, globally invertible rules such as $\chi$ seem a better option, although a CA cannot implement their inverse.
	
	More recently, Grassi et al.~\cite{grassi22} generalized the study of $\chi$-like mappings to non-binary alphabets, with the motivation of developing symmetric primitives for secure multiparty computation (SMPC), zero-knowledge proofs (ZKP) and fully homomorphic encryption schemes (FHE). The main result proved by the authors is that PBCA (there called \emph{liftings}) defined by quadratic rules of diameter $d=2$ and $d=3$ over the finite field $\F_p$ (for a prime $p\ge 3$) are never invertible. In follow-up work, Giordani et al.~\cite{giordani23} overcame this problem by considering \emph{non-uniform} CA, i.e., CA, where each cell can use a different local rule to update its state. Again, the terminology of this paper is not aligned with the CA literature, as this type of mapping is called a shift-invariant lifting with multiple local maps instead. From a different angle, Grassi~\cite{grassi23} considered instead non-invertible mappings with bounded surjectivity since invertibility is not strictly required in SMPC, ZKP, or FHE applications. In particular, the author proved that the simple local rule $F: \F_p^2 \to \F_p$ of the form $F(x_0, x_1) = x_0^2 +x_1$ gives rise to a PBCA $F: \F_p^n \to \F_p^n$ that minimizes the probability of collisions, and it is $2^n$-bounded surjective. Daemen et al.~\cite{daemen24} analyzed the differential and linear propagation properties of this mapping, discovering the fascinating fact that they follow the same rules up to a relabeling of the digits.
	
	A related research trend concerns \emph{rotation symmetric S-boxes}, which are, in essence, PBCA where the diameter of the local rule equals the size of the CA lattice. Hence, the neighborhood of each cell corresponds to the entire CA input up to cyclic shifts. Rijmen et al.~\cite{rijmen} proved that bijective S-boxes defined as \emph{power maps} are linearly equivalent to rotation-symmetric S-boxes. This result is particularly interesting since S-boxes based many practical symmetric primitives employ power maps. Kavut~\cite{kavut} classified all $6\times 6$ bijective rotation-symmetric S-boxes up to affine equivalence, remarking that there exist only $4$ functions with the best possible trade-off of nonlinearity, differential uniformity, and algebraic degree. Later, Liu et al.~\cite{liu} provided a construction for rotation-symmetric S-boxes satisfying \emph{perfect diffusion}. 
	
	When the size increases beyond $6\times 6$, exhaustive enumeration of rotation-symmetric S-boxes becomes unfeasible, even by resorting to equivalence classes techniques as done by Kavut. Similarly to the metaheuristics-based search for local rules with good cryptographic properties, a research trend took hold for the search of rotation-symmetric S-boxes. Picek et al.~\cite{picek17a} explored the use of Genetic Programming (GP) to optimize rotation-symmetric S-boxes with sizes between $5\times 5$ to $8\times 8$, achieving optimal values of nonlinearity and differential uniformity up to $7 \times 7$. Then, in follow-up research, Picek et al.~\cite{picek17} further investigated this GP optimization strategy by considering implementation criteria besides the cryptographic properties. The authors obtained rotation-symmetric S-boxes with optimal nonlinearity, differential uniformity, and low implementation costs comparable to those of other S-boxes in the state of the art. Finally, Mariot et al.~\cite{mariot19} proved that the best bounds on nonlinearity and differential uniformity for CA-based S-boxes correspond to the rotation-symmetric case (i.e., diameter equal to the CA size). Moreover, the authors used GP to reverse-engineer an S-box by finding the shortest CA rule that synthesizes it.
	
	\subsection{Shortcomings and Insights}
	\label{subsec:ins-block}
	The survey in the previous section shows that the CA approach is extensively used also by cryptographers to design block ciphers, although the terminology may vary. Nonetheless, one can still see a sharp methodological difference between the papers published in CA/natural computing venues and those published in cryptography conferences and journals. In particular, it is possible to identify a third shortcoming of the former ones:
	\begin{shortcoming}
		\label{shc:third}
		Using non-standard paradigms to design block ciphers, such as iterating CA as dynamical systems, hinders the security analysis. A general appeal to the confusion and diffusion principles, either by similarity or metaphor, is not a sound approach.
	\end{shortcoming}
	The issue here is quite similar to Shortcoming~\ref{shc:second} for the design of CA-based PRG with non-relevant models of pseudorandom generators. Most of the properties of S-boxes, such as nonlinearity and differential uniformity, are linked to attacks against specific paradigms of block ciphers, such as the Substitution-Permutation Network or components thereof. However, as remarked in Section~\ref{subsec:iter}, several works in the iterating CA design paradigm make security claims either based on empirical tests or by considering properties that have not been proven relevant for this model of block cipher. Thus, the following insight is relatively straightforward:
	\begin{insight}
		\label{ins:third}
		Confusion and diffusion, as formulated by Shannon, are abstract properties. As such, they need to be translated into practical design principles. It is preferable to work with well-established design paradigms for block ciphers (e.g., SPN ciphers and sponges) and insert CA as building blocks inside those	paradigms (e.g., as S-boxes).
	\end{insight}
	It is always possible to propose a CA-based design paradigm for block ciphers, but a lot more effort is required to vet its security: in practice, this	would entail performing rigorous linear and differential cryptanalysis and providing bounds for the best attacks. Hence, the recommendation to work with well-known paradigms is merely a matter of convenience since their security is better understood. Moreover, as illustrated in Section~\ref{subsec:single}, there are plenty of directions for future research on low-level primitives based on CA to be plugged into known symmetric designs. An interesting open problem here is to verify or refute Grassi et al.'s conjecture in~\cite{grassi22}, which states that there exists a finite integer $n_{max}(m)$ such that a PBCA $F: \F_p^n \to \F_p^n$ based on a quadratic rule of diameter $m$ is never invertible for all $n \ge n_{max}(m)$.
	
	Upon closer inspection, there is even a more basic reason why it is not a good idea to base the design of a block cipher entirely on CA, which explains why cryptographers only use them as sub-components for the confusion layer. The following last shortcoming summarizes this aspect: 
	\begin{shortcoming}
		\label{shc:fourth}
		CA are simply bad for diffusion.
	\end{shortcoming}
	Indeed, ciphertext differences cannot spread arbitrarily fast in a CA since the local rule's diameter binds their propagation speed. Hence,
	using only CA for the diffusion layer of a cipher usually entails the iteration of the CA for multiple time steps, as done in Gutowitz's proposal~\cite{gutowitz93}. However, this approach is not ideal for efficiency reasons: why should a cryptographer prefer a CA-based diffusion layer, which requires several time steps (and thus, clock cycles) when there are non-local methods that allow to achieve the same effect in a \emph{single step}? This leads the discussion to a concluding insight:
	\begin{insight}
		\label{ins:fourth}
		For certain components of a block cipher, it is better to abandon the CA approach. Non-local transformations are usually better, especially for the diffusion phase.
	\end{insight}
	Alternatively, a possible direction for future research would be to investigate how to implement CA-based diffusion layers with the minimal number of steps possible and compare it to the best non-local transformations in the state of the art. An interesting idea here is to leverage the combinatorial designs perspective on CA adopted in some works such as~\cite{mgfl-desi-2020,gadouleau23}. Specifically, a CA-based construction for generic orthogonal arrays could be employed to obtain MDS matrices, which are widely used in designing optimal diffusion layers (the best-known example being the diffusion layer of AES~\cite{daemen20}).
	
	\section{An Outlook on Future Research}
	\label{sec:outro}
	This paper gave a partial overview of the vast field of CA-based cryptography, narrowing the attention to the use cases of stream and block ciphers. As a result, this short survey identified four shortcomings in the literature that addresses the design of CA-based symmetric primitives mainly from the CA standpoint: over-reliance on empirical and statistical tests to make security claims, misalignment between the PRG models studied in cryptography and Wolfram's PRG model, adoption of non-standard paradigms for block ciphers, and poor diffusion inherent to the CA model. Accordingly, the paper also formulated four corresponding insights to mitigate such shortcomings, looking at how the cryptography literature analyzes the security properties of stream and block ciphers.
	
	The discussion also emphasized that cryptographers extensively use CA in their works. However, the difference in the terminology could explain the little permeability between the CA and cryptography research communities. The hope is that this paper will help to bridge these two communities, as they often work on closely related problems, and there are broad avenues for future collaborations.  
	
	\subsubsection*{Acknowledgements.} This research is partially supported by the PRIN 2022 PNRR project "Cellular Automata Synthesis for Cryptography Applications (CASCA)" (P2022MPFRT) financed by the European Union–-Next Generation~EU.
	
	\bibliographystyle{splncs04}
	\bibliography{references}
	
\end{document}